\documentclass[sigconf]{acmart}

\AtBeginDocument{%
  }

\copyrightyear{2026}
\acmYear{2026}
\setcopyright{cc}
\setcctype{by}
\acmConference[MSR '26]{23rd International Conference on Mining Software Repositories}{April 13--14, 2026}{Rio de Janeiro, Brazil}
\acmBooktitle{23rd International Conference on Mining Software Repositories (MSR '26), April 13--14, 2026, Rio de Janeiro, Brazil}
\acmDOI{10.1145/3793302.3793583}
\acmISBN{979-8-4007-2474-9/2026/04}

\usepackage{booktabs}
\usepackage{multirow}
\usepackage{xcolor}
\usepackage{balance}
\usepackage{hyperref}
\usepackage{colortbl}

\definecolor{ored}{HTML}{FB8957}
\definecolor{dkblue}{HTML}{8AAFC0}

\newsavebox{\takeawaysavebox}

\newenvironment{takeawaybox}[1][]{%
  \par\vspace{2pt}\noindent
  \setlength{\fboxsep}{2pt}%
  \begin{lrbox}{\takeawaysavebox}%
  \begin{minipage}{\dimexpr\linewidth-2\fboxsep\relax}%
  \ifx\relax#1\relax\else
    {\textbf{#1}}\par\vspace{2pt}%
  \fi
}{%
  \end{minipage}%
  \end{lrbox}%
  \colorbox{dkblue!50}{\usebox{\takeawaysavebox}}%
  \par\vspace{2pt}%
}

\begin{document}

\title{Comparing AI Coding Agents: A Task-Stratified Analysis of Pull Request Acceptance}

\author{Giovanni Pinna}
\orcid{0000-0001-8268-3447}
\affiliation{%
  \institution{University of Trieste}
  \city{Trieste}
  \country{Italy}
}
\email{giovanni.pinna@phd.units.it}

\author{Jingzhi Gong}
\orcid{0000-0003-4551-0701}
\affiliation{%
  \institution{King's College London}
  \city{London}
  \country{UK}
}
\email{jingzhi.gong@kcl.ac.uk}

\author{David Williams}
\orcid{0009-0004-9828-2639}
\affiliation{%
  \institution{University College London}
  \city{London}
  \country{United Kingdom}
}
\email{david.williams.22@ucl.ac.uk}

\author{Federica Sarro}
\orcid{0000-0002-9146-442X}
\affiliation{%
  \institution{University College London}
  \city{London}
  \country{United Kingdom}
}
\email{f.sarro@ucl.ac.uk}

\renewcommand{\shortauthors}{Pinna et al.}

\begin{abstract}
The rapid adoption of AI-powered coding assistants is transforming software development practices, yet systematic comparisons of their effectiveness across different task types and over time remain limited. This paper presents an empirical study comparing five popular agents (OpenAI Codex, GitHub Copilot, Devin, Cursor, and Claude Code), analyzing 7,156 pull requests (PRs) from the AIDev dataset. Temporal trend analysis reveals heterogeneous evolution patterns: Devin exhibits the only consistent positive trend in acceptance rate (+0.77\% per week over 32 weeks), whereas other agents remain largely stable. Our analysis suggests that the PR task type is a dominant factor influencing acceptance rates: documentation tasks achieve 82.1\% acceptance compared to 66.1\% for new features---a 16 percentage point gap that exceeds typical inter-agent variance for most tasks. OpenAI Codex achieves consistently high acceptance rates across all nine task categories (59.6\%–88.6\%), with stratified Chi-square tests confirming statistically significant advantages over other agents in several task categories. However, no single agent performs best across all task types: Claude Code leads in documentation (92.3\%) and features (72.6\%), while Cursor excels in fix tasks (80.4\%).
\end{abstract}


\maketitle

\section{Introduction}
The Software Engineering (SE) landscape has undergone a significant transformation with the emergence of AI-powered coding assistants~\cite{fan2023llmse, hou2023llmsurvey, gong2025ga4gc}.
Tools such as GitHub Copilot~\cite{ziegler2022productivity}, OpenAI Codex~\cite{chen2021codex}, and autonomous agents like Devin have moved beyond simple code completion to generating entire functions, fixing bugs, and creating pull requests autonomously~\cite{cognition2024devin}.
\citet{li2025rise} characterize this transition as ``Software Engineering 3.0,'' defined by autonomous AI teammates capable of completing complex development activities with minimal human supervision.

This phenomenon raises several fundamental questions: \textit{How do different AI coding agents compare when deployed in real-world software development workflows? What factors influence their effectiveness? Do outcome metrics vary across agents, task types, and time?}
These questions have implications for three key stakeholder groups: practitioners selecting tools for their teams; tool developers prioritizing capability improvements; and researchers designing robust evaluation methodologies for AI-assisted software engineering.

Moreover, open challenges remain regarding methodological shortcomings in LLM performance evaluations~\cite{williams2026reflection}, including the fact that global performance metrics may fail to distinguish between agent capability and task assignment. For instance, an agent that handles mostly documentation tasks (which tend to have higher acceptance rates according to our results) may appear superior to one that handles complex feature implementations, even if the latter has stronger underlying capabilities.
This confounding effect has been documented in past observational studies of SE practices~\cite{kalliamvakou2014promises}, where uncontrolled variables can lead to misleading conclusions.

In this paper, we tackle some of the aforementioned questions by examining temporal trends to understand how AI coding agents perform over extended deployment periods, remaining agnostic about underlying mechanisms such as model updates, user behavior changes, or task distribution shifts.
Additionally, we address workload heterogeneity through \emph{task-stratified comparisons}, evaluating agents within each task category (e.g., feature, fix, documentation) to control for confounding factors in different task distributions. Specifically, we answer the following three research questions by analyzing the AIDev dataset~\cite{aidev2025, li2025aidev}:

\noindent \textbf{RQ1: Does AI coding agent performance evolve over time?} 
We investigate whether agents demonstrate measurable changes in acceptance rates over extended observation periods.

\noindent \textbf{RQ2: What factors are associated with performance?} 
We examine how task type, review frequency, and other factors correlate with acceptance rates, and whether differences in task distribution across agents confound global comparisons.

\noindent \textbf{RQ3: How do different agents compare?} 
We perform task-stratified statistical comparisons of five major AI coding agents to identify differences in performance.

Our contributions include: (i) temporal analysis revealing heterogeneous evolution patterns across agents, with Devin showing the only consistent improvement; (ii) a task-stratified comparison methodology that controls for workload heterogeneity when evaluating AI agents; (iii) empirical evidence that task type is a dominant performance factor, with a 29 percentage point (pp) gap between task types; and (iv) analysis of agent-specific strengths across task types, showing that no single agent outperforms others in all tasks.

\section{Background and Related Work}
The technological evolution of LLM-based development tools has been rapid.
OpenAI Codex launched in August 2021~\cite{chen2021codex}, powering GitHub Copilot, which reached general availability in June 2022~\cite{github2022copilot}. 
Since then, we have witnessed the advent of autonomous agents: Devin by Cognition AI (March 2024) achieved 13.86\% on SWE-bench unassisted, a substantial improvement over the prior 1.96\% state-of-the-art~\cite{cognition2024devin}; Cursor grew to over one million users by 2025~\cite{cursor2025}; and Claude Code became widely available in May 2025 as Anthropic's terminal-based agentic coding tool~\cite{anthropic2025claude}.
Research on productivity impacts presents mixed findings: \citet{peng2023impact} found Copilot users completed tasks 55.8\% faster, yet METR~\cite{metr2025} observed experienced developers were 19\% slower, despite believing otherwise.
\citet{yetistiren2023copilot} found that only 28.7\% of Copilot suggestions were directly usable, with quality varying by language and task complexity. 
Security vulnerabilities in AI-assisted code have also been documented~\cite{perry2023users}.
Finally, \citet{he2025does} conducted a longitudinal study of Cursor adoption, finding that velocity gains were concentrated in the first two months before returning to baseline, while technical debt---measured through static analysis warnings and code complexity---accumulated persistently. 
These findings reinforce concerns about using acceptance rate as a sole quality proxy, as short-term acceptance may mask long-term maintenance burdens.

\section{Methodology}
\subsection{Dataset and Preprocessing}\label{sec:dataset}

We utilize the AIDev dataset~\cite{aidev2025}, which contains AI-generated pull requests from GitHub repositories with at least 100+ stars (the AIDev-POP subset). 
The raw dataset contains 33,596 PRs, along with corresponding file-level changes, reviews, comments, and LLM-classified task categories.
From this data, we derive a filtered dataset of 7,156 PRs by applying the following quality criteria:
We retain \textcircled{1} \emph{closed} PRs from \textcircled{2} repositories with \emph{permissive licenses} (MIT or Apache-2.0) and \textcircled{3} required that each PR \emph{receive at least one review or comment from someone other than the creator before closure} (ensuring meaningful evaluation).
Before analysis, we aggregate commit- and review-level data to the PR level, computing additions, deletions, files changed, commits, reviews, and comments per PR.
For temporal analysis, we group PRs by the week of their creation date; an \emph{active week} is any week containing at least one PR from a given agent.
Table~\ref{tab:agents} summarizes the resulting agent distribution. 
The ``Start'' column reflects dataset coverage rather than official releases; ``Weeks'' reports the number of active weeks containing at least one PR.
Cursor's 13 active weeks span May--August 2025, representing a concentrated adoption period compared to Devin's extended 32-week observation window.
These temporal differences have two implications: agents cannot be compared across equivalent maturity periods, and observed patterns may reflect deployment contexts rather than intrinsic capabilities.
We address this limitation through sensitivity analysis using aligned time windows in Section~\ref{sec:discussion}.

\begin{table}[tb]
\centering
\caption{Agent PR distributions, active weeks, and overall acceptance rates in the filtered dataset.}
\vspace{-0.3cm}
\label{tab:agents}
\resizebox{2.5in}{!}{
\begin{tabular}{lrrrrr}
\toprule
\textbf{Agent} & \textbf{Start} & \textbf{PRs} & \textbf{Weeks} & \textbf{PRs/Wk} & \textbf{Acc. Rate} \\
\midrule
Devin & 12/24/24 & 2,252 & 32 & 70.4 & 61.6\% \\
OpenAI Codex & 05/16/25 & 2,002 & 12 & 166.8 & 77.9\%\\
GitHub Copilot & 05/19/25 & 2,194 & 11 & 199.5 & 68.0\% \\
Cursor & 05/01/25 & 569 & 13 & 43.8 & 74.5\% \\
Claude Code & 02/24/25 & 139 & 19 & 7.3 & 71.9\% \\
\midrule
\textbf{Total} & & \textbf{7,156} & \textbf{87} & \textbf{---} & \textbf{69.3\%} \\
\bottomrule
\end{tabular}
}
\end{table}

\subsection{Metrics and Statistical Methods}

We characterize success outcomes through \emph{acceptance rate}, defined as the proportion of merged PRs (those with a non-null \texttt{merged\_at} timestamp).
Since our filtering retains only closed PRs (Section~\ref{sec:dataset}), the acceptance and rejection rates sum to 100\% in our analysis.
We define \emph{task-stratified observations} as unique (agent, task type, week) combinations where at least one PR was submitted.
Summing across all agents and the 12 task types yields 552 total task-stratified observations (Table~\ref{tab:tasks}).
To address RQ1, we measure \emph{temporal trends} in acceptance rate per agent by fitting linear regression models defined as $y = \beta_0 + \beta_1 \cdot t + \varepsilon$, where $\beta_1$ represents the weekly rate of change.
A slope of $+0.01$ would indicate a one-percentage-point weekly increase.
We report $R^2$ values to characterize model fit, with values closer to 1 indicating stronger linear trends and values near 0 suggesting no consistent temporal pattern.
To capture non-linear patterns, we apply LOESS (Locally Estimated Scatterplot Smoothing) with a fraction parameter $frac=0.5$~\cite{cleveland1979robust}, balancing smoothness against sensitivity to local trends.
For RQ2, we examine task-level acceptance rates and \emph{review frequency}, defined as the number of reviews per PR (indicating the extent of evaluation each agentic contribution undergoes).
For RQ3, we conduct task-stratified pairwise agent comparisons using Pearson's Chi-square tests of independence~\cite{pearson1900criterion}.
To avoid over-interpreting sparse observations, we exclude comparisons where at least one agent has fewer than 10 total PRs for a given task. We additionally exclude three task types (\texttt{style}, \texttt{revert}, and \texttt{other}) because multiple agents have zero observations for these categories in our filtered dataset. With these criteria, comparisons across the 10 possible agent pairs and 9 remaining task types yield 64 task-stratified tests.
For comparisons where expected cell frequencies fall below 5, we apply Fisher's exact test~\cite{fisher1922interpretation} (17 of 64 comparisons).
To control for multiple comparisons, we apply the Bonferroni correction across the executed test set, yielding an adjusted significance threshold of $\alpha = 0.05/64 \approx 0.00078$.
In addition, we report the phi coefficient ($\phi$)~\cite{cohen1988statistical} to measure effect size, following standard interpretation thresholds: $\phi < 0.1$ (negligible), $0.1 \leq \phi < 0.3$ (small), $0.3 \leq \phi < 0.5$ (medium), and $\phi \geq 0.5$ (large).

Finally, we emphasize that this study is observational: we document patterns without claiming causality, acknowledging potential confounders such as repository concentration, user population dynamics, model updates, and shifts in task distributions.

\section{Results}

\subsection{RQ1: Performance Over Time}

\begin{figure}[tb]
    \centering
    \includegraphics[width=0.9\columnwidth]{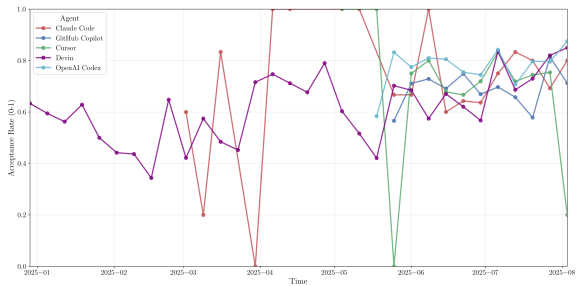}
    \vspace{-0.3cm}
    \caption{RQ1. Acceptance rate over time per agent.}
    \vspace{-0.3cm}
    \label{fig:learning_curves}
\end{figure}

Figure~\ref{fig:learning_curves} illustrates heterogeneous performance patterns across agents. Devin exhibits a positive trend in acceptance rate (+0.77\% per week, $R^2 = 0.34$), increasing from approximately 60\% to 80\%.
Devin's LOESS curve suggests three phases: an initial phase ($\sim$60\%), an increasing phase (to $\sim$80\%), and a stabilization phase---though we cannot isolate underlying causal mechanisms.
Notably, while Devin shows improvement, its weekly variance remains high, suggesting inconsistent performance despite the positive trend.
In contrast, OpenAI Codex and GitHub Copilot exhibit plateau behavior, with consistently high acceptance rates from their first observed week.
\begin{takeawaybox}
\textbf{RQ1:} Devin is the only agent showing consistent performance improvement over time, while other agents maintain stable acceptance rates.
\end{takeawaybox}

\subsection{RQ2: Factors Associated with Performance}
Table~\ref{tab:tasks} reports acceptance rates calculated at the task level (i.e., merged PRs divided by total PRs for each task type).
From these results, task type emerges as a dominant factor in agentic PR outcomes.
Table~\ref{tab:tasks} shows a 29pp gap between the highest (\texttt{chore}: 84.0\%) and lowest (\texttt{perf}: 55.4\%) acceptance rates.
Among high-volume task types, documentation achieves 82.1\% versus 66.1\% for features---a 16 percentage-point gap.
This pattern suggests that AI agents achieve higher acceptance on well-structured tasks, though alternative explanations exist (e.g., less stringent review standards for documentation PRs).
\begin{table}[tb]
\centering
\caption{RQ2. Acceptance rate by task type. MAR: Mean acceptance rate; SD: standard deviation; Obs: \# weekly agent-task observations.}
\vspace{-0.3cm}
\label{tab:tasks}
\resizebox{\columnwidth}{!}{
\begin{tabular}{lrrrrrrrrrrrr}
\toprule
  & \textbf{chore} & \textbf{docs} & \textbf{style} & \textbf{ci} & \textbf{build} & \textbf{refactor} & \textbf{feat} & \textbf{fix} & \textbf{test} & \textbf{revert} & \textbf{perf} & \textbf{other} \\
\midrule
\textbf{MAR} & \textbf{84.0\%} & 82.1\% & 78.1\% & 75.0\% & 72.5\% & 71.2\% & 66.1\% & 66.0\% & 61.5\% & 60.0\% & 55.4\% & 0.0\% \\
\textbf{SD} & 36.7\% & 38.4\% & 42.0\% & 43.5\% & 44.8\% & 45.3\% & 47.3\% & 47.4\% & 48.8\% & 54.8\% & 50.2\% & 0.0\% \\
\textbf{Obs} & 58 & 71 & 25 & 41 & 40 & 69 & 81 & 77 & 53 & 4 & 29 & 4 \\
\bottomrule
\end{tabular}
}
\end{table}
Beyond task type, we observe variation in review frequency across agents. 
GitHub Copilot PRs receive substantially more reviews (4.94/PR) than OpenAI Codex PRs (1.39/PR), coinciding with lower acceptance rates (68.0\% vs. 77.9\%). 
However, this does not imply a causal link: varying review counts may reflect confounders such as task complexity or repository reviewing policies.
\looseness=-1
\begin{takeawaybox}
\textbf{RQ2:} PR task type is a dominant factor---the 29\% acceptance rate gap between task categories substantially exceeds inter-agent variance within most categories.
\end{takeawaybox}

\subsection{RQ3: Task-Stratified Agent Comparison}
\begin{figure}[tb]
    \centering
    \includegraphics[width=0.9\columnwidth]{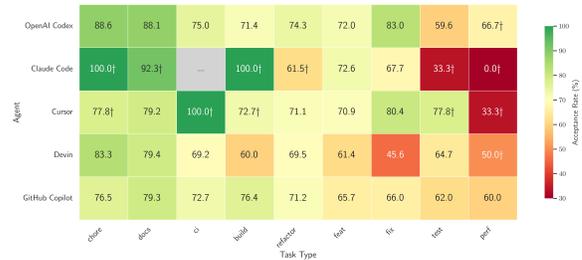}
    \vspace{-0.3cm}
    \caption{RQ3. Acceptance rates (\%) by agent and task type. ``$\dagger$'' indicates $<$20 PRs; ``---'' indicates zero PRs.}
    \label{fig:heatmap}
\end{figure}

Global comparisons may confound agent capability with task distribution. 
Table~\ref{tab:task_dist} reveals substantial differences in task distributions across agents: GitHub Copilot handles proportionally more \texttt{fix} tasks (41.6\%) than other agents, while Claude Code focuses heavily on \texttt{feat} tasks (52.5\%).
It should be noted that any results for Claude Code must be interpreted with caution due to the limited sample size (139 PRs total).
If one agent handles mostly documentation (82\% acceptance rate) while another handles fixes (66\% acceptance), global metrics are misleading.
Therefore, we stratify by task type.
\begin{table}[tb]
\centering
\caption{Task distribution by agent (\% of each agent's PRs).  $\dagger$ indicates $<$20 PRs.}
\vspace{-0.3cm}
\label{tab:task_dist}
\resizebox{2.5in}{!}{
\begin{tabular}{lrrrr}
\toprule
\textbf{Agent} & \textbf{feat} & \textbf{fix} & \textbf{docs} & \textbf{refactor} \\
\midrule
OpenAI Codex & 28.9\% & 28.3\% & 11.3\% & 7.0\% \\
GitHub Copilot & 29.2\% & 41.6\% & 8.6\% & 6.7\% \\
Devin & 41.5\% & 25.7\% & 11.4\% & 8.3\% \\
Cursor & 33.2\% & 25.1\% & 16.9\% & 6.7\% \\
Claude Code & 52.5\% & 22.3\% & 9.4\%$\dagger$ & 9.4\%$\dagger$ \\
\bottomrule
\end{tabular}
}
\end{table}
Figure~\ref{fig:heatmap} visualizes acceptance rates for each agent-task combination across nine categories (\texttt{style}, \texttt{revert}, \texttt{other} were excluded due to insufficient data).
Cells marked ``$\dagger$'' contain $<$20 PRs; the cell marked ``---'' indicates zero PRs (Claude Code has no \texttt{ci} PRs in our filtered dataset).
OpenAI Codex achieves consistently high acceptance rates across the nine task categories, ranging from 59.6\% (\texttt{test}) to 88.6\% (\texttt{chore}), and leads in \texttt{fix} (83.0\%) and \texttt{refactor} (74.3\%) tasks.
However, other agents achieve higher rates in specific categories: Claude Code leads in \texttt{docs} (92.3\%, albeit with few samples) and \texttt{feat} (72.6\%), while Cursor excels in \texttt{test} (77.8\%, also with few samples).
This task-dependent variation suggests that no single agent outperforms others across all categories, though unobserved confounders such as user expertise or repository characteristics remain uncontrolled.
This pattern is illustrated in \texttt{test} tasks, where Devin achieves a higher acceptance rate (64.7\%) than OpenAI Codex (59.6\%), despite exhibiting substantially lower global performance (61.6\% vs. 77.9\%, shown in Table~\ref{tab:agents}).
We further examine whether task distributions shift over time within agents in Section~\ref{sec:discussion}.
Table~\ref{tab:chisquare} presents task-stratified Chi-square tests of independence; only results passing the Bonferroni-adjusted significance threshold ($\alpha = 0.05/64 \approx 0.00078$) are shown, with full results available in the replication package~\cite{pinna_comparing_ai}.
Of the 64 stratified comparisons across all 10 agent pairs, 6 are significant, with the strongest effect observed in \texttt{fix} tasks (Codex vs. Devin: $\phi = 0.39$, medium effect).
All significant comparisons involve \texttt{fix} (5 of 6) or \texttt{feat} (1 of 6) tasks, suggesting that performance differences between agents are most detectable in these core development activities.
Notably, comparisons involving Devin on \texttt{fix} tasks show consistent patterns: Devin underperforms relative to Codex ($\phi = 0.39$), Copilot ($\phi = 0.20$), and Cursor ($\phi = 0.28$), indicating a potential weakness in bug-fixing capabilities (which can also be observed in terms of raw acceptance rates in Table~\ref{fig:heatmap}).
However, it must be noted that without controlling for other possible confounders, such as repository-level effects, these findings remain observational.

\begin{table}[tb]
\centering
\caption{RQ3. Pairwise agent comparisons. Only Bonferroni-significant results are shown ($\alpha = 0.05/64 \approx 0.00078$).}
\vspace{-0.3cm}
\label{tab:chisquare}
\resizebox{2.5in}{!}{
\begin{tabular}{llrlrl}
\toprule
\textbf{Comparison} & \textbf{Task} & \textbf{p-value} & \textbf{Winner} & \textbf{$\phi$} & \textbf{Effect} \\
\midrule
Codex vs Devin & \texttt{fix} & $<$0.0007 & Codex & 0.39 & medium \\
Copilot vs Devin & \texttt{fix} & $<$0.0007 & Copilot & 0.20 & small \\
Devin vs Cursor & \texttt{fix} & $<$0.0007 & Cursor & 0.28 & small \\
Codex vs Copilot & \texttt{fix} & $<$0.0007 & Codex & 0.19 & small \\
Codex vs Devin & \texttt{feat} & $<$0.0007 & Codex & 0.11 & small \\
Cursor vs Copilot & \texttt{fix} & $<$0.0007 & Cursor & 0.11 & small \\
\bottomrule
\end{tabular}
}
\end{table}
\begin{takeawaybox}
\textbf{RQ3:} No single agent consistently outperforms all others; OpenAI Codex shows consistent strength across categories, but Claude Code and Cursor lead in specific task types.
\end{takeawaybox}

\section{Discussion}
\label{sec:discussion}

Our results reveal that AI agents exhibit heterogeneous temporal trends: Devin's acceptance rate increases over time, while others remain relatively stable. 
OpenAI Codex's consistently high acceptance rates (60--88\% across all tasks), leading in \texttt{fix} and \texttt{refactor} tasks, suggest strong underlying capabilities. However, Claude Code and Cursor outperform Codex in specific tasks.
The 44.4 percentage-point gap on \texttt{test} tasks (Cursor 77.8\% vs. Claude Code 33.3\%) represents the largest inter-agent disparity across task categories, suggesting that more complex tasks exhibit greater differences.

\textbf{Sensitivity Analysis: Aligned Time Windows.}
To address concerns about temporal comparability, we repeated our analysis using only the 11-week window common to all agents (May 19--July 30, 2025).
Results remain consistent: OpenAI Codex has the highest acceptance rate (79.9\%), followed by Cursor (74.4\%), Claude Code (72.6\%), Devin (68.0\%), and GitHub Copilot (68.0\%). 
Claude Code shows moderate variance (std=11.7\%) with 113 PRs in this window.

\textbf{Task Distribution Heterogeneity.}
As shown in Table~\ref{tab:task_dist}, task distributions vary substantially across agents, underscoring the importance of task-stratified comparisons. We further examined whether task type distributions shift over time within agents, as such compositional changes could confound the interpretation of performance trends. 
Comparing early versus late observation periods, we find varying patterns: Devin and GitHub Copilot shifted toward \texttt{feat} tasks ($+$9.8pp and $+$8.1pp), while Codex shifted away ($-$5.7pp), and Claude Code shows the largest compositional change ($+$18.4pp \texttt{feat}, $-$23.0pp \texttt{docs}).
These shifts have competing implications for trend interpretation: Devin's acceptance rate increased despite handling a progressively larger proportion of more complex tasks (+9.8pp shift toward \texttt{feat} tasks).
This shift toward harder tasks suggests the observed improvement may understate actual capability gains.

\textbf{Implications for Practitioners.} Our results indicate that agent choice matters most for bug fixes (Cursor 80.4\%, Codex 83.0\% vs. Devin 45.6\%) and testing tasks (33--78\% range); documentation shows minimal differentiation ($>$79\%). Depending on the use case, practitioners may want to adopt a combination of agents to address different workflows.

\textbf{Implications for Researchers.} Task stratification should become standard practice: future studies should report task distributions alongside global metrics, stratify comparisons by task type, and flag underrepresented types. Given the limitations of acceptance rates, complementary metrics (static analysis, complexity, maintenance burden) are needed~\cite{he2025does}. Extended discussion of practical implications and actionable recommendations is available in the \href{https://github.com/giovannipinna96/Comparing_AI_Coding_Agents}{\textcolor{blue}{supplementary material}}~\cite{pinna_comparing_ai}.

\textbf{Threats to Validity.} We cannot establish causality (internal validity)---observed trends could result from model updates, user learning, or task distribution shifts. 
Our data covers only 8 months from 100+ star repositories (external validity), limiting generalization to smaller projects. 
PR acceptance does not equal code quality (construct validity), as merged PRs may contain bugs~\cite{perry2023users}. 
Additionally, repository-level clustering---where multiple PRs originate from the same repository---is not explicitly modeled in our Chi-square tests.
If certain repositories systematically accept or reject PRs regardless of quality, this could inflate statistical significance; future work should consider mixed-effects models to account for repository-level variance.
Observation windows vary substantially (11--32 active weeks), and some agents show discontinuous activity, requiring cautious interpretation of observed trends. 
Finally, agent representation ranges from 139 to 2,252 PRs, with Claude Code's sparse data limiting reliability.

\textbf{Ethical Considerations.} This study uses only publicly available data from repositories with permissive licenses (Apache-2.0 and MIT)~\cite{MSR20GoldKrinke}. Through our analysis, no attempt is made to re-identify any individuals present in the dataset, and only aggregate findings are reported. The study received ethics approval from University College London (LREC no.: 2568).

\section{Conclusion}
Analyzing 7,156 pull requests from five AI coding agents, we find that \textbf{task type is a dominant factor over} acceptance rates (29pp gap between types), motivating our \textbf{task-stratified comparison} methodology.
Our stratified analysis reveals that no single agent performs best across all task types.
OpenAI Codex achieves consistently high acceptance rates (59.6\%--88.6\%), but Claude Code leads in documentation (92.3\%) and features (72.6\%), while Cursor excels in fix tasks (80.4\%).
These task-dependent variations provide evidence against simple ``best agent'' rankings.
Temporal analysis reveals heterogeneous evolution patterns: Devin exhibits the only consistent positive trend (+0.77\%/week over 32 weeks), while other agents maintain stable acceptance rates.
These findings underscore the importance of considering both task context and temporal dynamics when evaluating AI coding agents.
Future work should incorporate code quality metrics and static analysis warnings to complement acceptance rate analysis.

\vspace{0.2cm}
\noindent \textbf{Data Availability.}
All data, scripts, results, and other supplementary materials are available on \href{https://github.com/giovannipinna96/Comparing_AI_Coding_Agents}{\textcolor{blue}{GitHub}} \cite{pinna_comparing_ai}.

\newpage

\balance
\bibliographystyle{ACM-Reference-Format}
\bibliography{ref}

\end{document}